\documentclass[twocolumn,trackchanges,twocolappendix]{aastex631}
\usepackage{natbib}
\usepackage{color}
\usepackage{graphicx}
\usepackage{multirow}

\newcommand{\NaEbv}{0.145}
\newcommand{\per }{3.2254367} 
\newcommand{\perErr }{0.0000003}
\newcommand{\vga }{280.7 }
\newcommand{\vgaErr }{0.4}
\newcommand{\sma }{30.48}
\newcommand{\smaErr }{0.07}
\newcommand{\daop }{-0.005}
\newcommand{\daopErr }{0.004}
\newcommand{\aop }{0.3} 
\newcommand{\aopErr }{1.0} 
\newcommand{\ecc }{0.0006}
\newcommand{\eccErr }{0.0003}
\newcommand{\incl }{87.43}
\newcommand{\inclErr }{0.21}
\newcommand{\q }{0.919 }
\newcommand{\qErr }{0.007 }

\newcommand{\avgEbvErr}{0.16 $\pm$ 0.02}
\newcommand{\vrotBFPri}{ 70 $\pm$ 7 }  
\newcommand{\vrotBFSec}{ 80 $\pm$ 8 }

\newcommand{\mPri}{19.02}
\newcommand{\mPriErr }{0.12}
\newcommand{\rPri }{7.70}
\newcommand{\rPriErr }{0.05}
\newcommand{\loggPri }{3.944}
\newcommand{\loggPriErr }{0.005}
\newcommand{\logLPri }{4.854}  

\newcommand{\teffPri }{34250} 
\newcommand{\teffPriErr }{500}
\newcommand{\mVPri }{14.963}    
\newcommand{\mVPriErr }{0.021}
\newcommand{\mIPri }{15.086}    
\newcommand{\mIPriErr }{0.021}
\newcommand{\mKPri }{15.316}    
\newcommand{\mKPriErr }{0.029}  

\newcommand{\mSec }{17.50} 
\newcommand{\mSecErr }{0.13} 
\newcommand{\rSec }{6.64} 
\newcommand{\rSecErr }{0.06}
\newcommand{\loggSec }{4.035}  
\newcommand{\loggSecErr }{0.007}
\newcommand{\logLSec }{4.700}  
   
\newcommand{\teffSec }{33750}   
\newcommand{\teffSecErr }{500}
\newcommand{\mVSec }{15.285}   
\newcommand{\mVSecErr }{0.025}
\newcommand{\mISec }{15.413}   
\newcommand{\mISecErr }{0.024}
\newcommand{\mKSec }{15.644}   
\newcommand{\mKSecErr }{0.029}

\newcommand{\VyE}[2]{#1 $\pm$ #2 }

\accepted{April 3, 2024}
\submitjournal{ApJ}

\shorttitle{Towards early-type systems as extragalactic milestones. III.}
\shortauthors{Taormina et al.}

\begin{document}

\title{Towards early-type eclipsing binaries as extragalactic milestones: III. Physical properties of the O-type eclipsing binary OGLE LMC-ECL-21568 in a quadruple system
\footnote{Based on observations collected at the European Southern Observatory, Chile}
\footnote{This paper includes data gathered with the 6.5m Magellan Clay Telescope at Las Campanas Observatory, Chile.}
} 

\correspondingauthor{M{\'o}nica Taormina}
\email{taormina@camk.edu.pl}

\author[0000-0002-1560-8620]{M{\'o}nica Taormina}
\affiliation{Centrum Astronomiczne im. Miko{\l}aja Kopernika PAN, Bartycka 18, 00-716 Warsaw, Poland}

\author{R.-P.  Kudritzki}
\affiliation{LMU M\"unchen, Universit\"atssternwarte, Scheinerstr. 1, 81679 M\"unchen, Germany}
\affiliation{Institute for Astronomy, University of Hawaii at Manoa, Honolulu, HI 96822, USA}

\author[0000-0003-3861-8124]{B. Pilecki}
\affiliation{Centrum Astronomiczne im. Miko{\l}aja Kopernika PAN, Bartycka 18, 00-716 Warsaw, Poland}

\author[0000-0002-9443-4138]{G. Pietrzy{\'n}ski}
\affiliation{Centrum Astronomiczne im. Miko{\l}aja Kopernika PAN, Bartycka 18, 00-716 Warsaw, Poland}

\author{I. B. Thompson}
\affiliation{Carnegie Observatories, 813 Santa Barbara Street, Pasadena, CA 91101-1292,  USA}

\author{J. Puls}
\affiliation{LMU M\"unchen, Universit\"atssternwarte, Scheinerstr. 1, 81679 M\"unchen, Germany}

\author[0000-0002-3125-9088]{M. G\'{o}rski}
\affiliation{Centrum Astronomiczne im. Miko{\l}aja Kopernika PAN, Bartycka 18, 00-716 Warsaw, Poland}
\author[0000-0003-1515-6107]{B. Zgirski}
\affiliation{Universidad de Concepci\'{o}n, Departamento de Astronom\'{i}a, Casilla 160-C, Concepci\'{o}n, Chile}

\author[0000-0002-7355-9775]{D. Graczyk}
\affiliation{Centrum Astronomiczne im.  Miko{\l}aja Kopernika PAN, Rabia{\'n}ska 8, 87-100 Toru{\'n}, Poland}

\author{W. Gieren}
\affiliation{Universidad de Concepci\'{o}n, Departamento de Astronom\'{i}a, Casilla 160-C, Concepci\'{o}n, Chile}

\author[0000-0003-0594-9138]{G. Hajdu}
\affiliation{Centrum Astronomiczne im. Miko{\l}aja Kopernika PAN, Bartycka 18, 00-716 Warsaw, Poland}

\begin{abstract}

We present the results from a complex study of an eclipsing O-type binary (Aa+Ab) with the orbital period $P_{A}=\per$ days, that forms part of a higher-order multiple system in a configuration (A+B)+C. 
We derived masses of the Aa+Ab binary $M_{1}= \mPri \pm \mPriErr \,M_\odot$, $M_{2}= \mSec \pm \mSecErr \,M_\odot$, radii $R_{1}= \rPri \pm \rPriErr \,R_\odot$, $R_{2}= \rSec \pm \rSecErr \,R_\odot$, and temperatures $T_1 = \teffPri \pm \teffPriErr$ K, $T_2 = \teffSec \pm \teffSecErr$~K.
From the analysis of radial velocities, we found a spectroscopic orbit of A in the outer A+B system with $P_{A+B}=195.8$ days ($P_{A+B}/P_{A}\approx 61$). In the O-C analysis, we confirmed this orbit and found another component orbiting the A+B system with $P_{AB+C}=2550$ days ($P_{AB+C}/P_{A+B}\approx 13$).
From the total mass of the inner binary and its outer orbit, we estimated the mass of the third object, $M_B \gtrsim 10.7 M_\odot$. From the light-travel time effect fit to the O-C data, we obtained the limit for the mass of the fourth component, $M_C \gtrsim 7.3 M_\odot$.  These extra components contribute to about 20\% to 30\% (increasing with wavelength) of the total system light.
From the comparison of model spectra with the multiband photometry, we derived a distance modulus of 18.59 $\pm$ 0.06 mag, a reddening of 0.16 $\pm$ 0.02 mag, and an $R_V$ of $3.2$.
This work is part of our ongoing project, which aims to calibrate the surface brightness-color relation for early-type stars.

\end{abstract}

\section{Introduction} \label{sec:intro}

OGLE LMC-ECL-21568 (henceforth BLMC-03) is located in the region known as the Lucke-Hodge 101 OB association \citep{lucke:1970} or NGC 2074 in the Large Magellanic Cloud (LMC). It was included in photographic, photometric, and spectroscopic studies of the region \citep[e.g.][]{westerlund:1961,testor:1998} and observed in various photometric surveys of the LMC \citep[e.g.][]{massey:2002b,zaritsky:2004}. 

In the first paper of a series, regarding the program to find eclipsing massive binaries in the LMC, \citet{massey:2012} presented a spectroscopic follow-up of BLMC-03 (under the ID [M2002] LMC 172231 from the catalog of \citealt{massey:2002b}) and classified it spectroscopically as O9V+O9.5V. They measured the radial velocities of the components and obtained an orbital solution. Combining these data with their new V-band photometry \citet{massey:2012} provided the first characterization of the system. They found the binary's light and radial velocity curves to be consistent with a circular orbit and assumed zero eccentricity for the rest of the study. We note that in their orbital solution, radial velocities show relatively high scatter compared to quoted errors. Moreover, the residuals for the two components are correlated, indicating a possible unaccounted additional variability in the system.
Finally, they found almost identical masses for the components, $17.5 \pm 0.3$ $M\odot$ and $17.6 \pm 0.3$ $M\odot$, and stellar radii of $7.0 \pm 0.3$ $R\odot$ and $6.5 \pm 0.3$ $R\odot$ for the primary and secondary star, respectively.

In our ongoing study of a carefully selected sample of early-type binaries in the LMC, we aim to obtain accurate and precise physical parameters of their components to calibrate the surface brightness-color (SBC) relation for hot stars. We have already analyzed two early-type systems in this galaxy  (\citealt{taormina:2019} - Paper I, \citealt{taormina:2020} - Paper II), and BLMC-03 is the next one on our list of the well-detached systems in the LMC.
 As we need the highest accuracy and precision possible and a light curve solution in the K-band, we decided to reanalyze this system using the currently available photometric data in visual and near-infrared passbands together with our new high-resolution spectra.

The paper is organized as follows. In Section \ref{sec:data} we present the data used in the study, in Section \ref{sec:analysis}, we describe the analysis and its direct results, and in Section \ref{sec:spectral} we present a detailed spectroscopic analysis. In Section \ref{sec:results}, we derive the final physical parameters of the binary components and orbital properties of the system. Our conclusions are presented in Section \ref{sec:conclusions}.

\begin{deluxetable}{lcc}
    \tabletypesize{\footnotesize}
    \tablecaption{General information about the system}\label{info}
    \tablehead{\colhead{parameter} & \colhead{unit} & \colhead{value}}
           \startdata
            our ID         & &  BLMC-03 \\
            OGLE ID        & &  LMC-ECL-21568 \\
            $\alpha_{2000}$ & [hh:mm:ss.ss]   &  05:38:58.08   \\ 
            $\delta_{2000}$  & [$\pm$dd:mm:ss.s]  &  -69:30:11.3  \\
            orbital period             &  [days] &  3.225450   \\
            $I_\mathrm{C}$ - band      &  [mag]  &  14.212 \\
            $V$ - band                 &  [mag]  &  14.125 \\
            $V-I_\mathrm{C}$          &  [mag]  &  -0.087 \\
            spectral type       &  &  O9V+O9.5V   \tablenotemark{a}
           \enddata
          \tablenotetext{a}{\cite{massey:2012}}          
\end{deluxetable}

\section{Observational Data}  \label{sec:data}

\subsection{Photometry}

As in our previous works (Paper I and II), we made use of the photometric data from the published catalogs of eclipsing binary stars in the LMC \citep{graczyk:2011,pawlak:2016}, based on phases 3 and 4 of the Optical Gravitational Lensing Experiment survey \citep[OGLE,][]{udalski:2003,udalski:2015}. We used in total 436 $I_\mathrm{C}$-band observations from the OGLE-III and 383 from the OGLE-IV survey as well as 45 and 72 V-band measurements from these two OGLE surveys, respectively. To have a better time coverage, we have also included in our analysis observations from the EROS-2 survey \citep{tisserand:2007}, using 244 data points in the EROS R-band, which is almost identical to the Johnson-Cousins $I_C$ filter. The light curves have been cleaned of outliers and any significant long-term brightness trends.

Near-infrared K-band data is essential for our goal of improving the calibration of the SBC relation.
For that reason, in the analysis of BLMC-03 we used 14 measurements from the VISTA Magellanic Clouds IR photometric survey \citep[VMC,][]{cioni:2011} together with 12 observations that we collected using the SOFI imaging camera on the 3.58m ESO New Technology Telescope \citep[NTT,][]{moorwood:1998} at the La Silla Observatory, Chile.
The latter were obtained using the Large Field setup with a field of view of 4.9' x 4.9' at a scale of 0.288" per pixel. The gain and readout noise were 5.4e/ADU and 0.4e, respectively. Deep J- and Ks-band observations of our target field were acquired under good seeing conditions during several nights between December 2015 and December 2018. 
To account for the frequent sky-level variations in the infrared spectral region, especially in the Ks-band, the observations were performed with a dithering technique. For the Ks and J-band observations, we took over 10 exposures\footnote{the exact number depended on the observing conditions} at a given pointing and moved the telescope to another position selected randomly within a 25" x 25" square. Between 15 and 25 such dithering positions were obtained for the Ks-band, and between 11 and 15 for the J-band.

To reduce the data we followed the procedure described in \cite{pietrzynski:2002}. The sky was subtracted from the images with a two-step process implying the masking of the stars with the {\tt xdimsum IRAF} package. The individual images for each field and filter were then flat-fielded and stacked into a final composite image. 
To obtain the light curve, the coordinates of all stars on separate images were crossmatched using {\tt DAOMATCH} and {\tt DAOMASTER} \citep{stetson:1994} programs.
The differential brightness of the eclipsing system was calculated by comparison with the selected sample of 13 comparison stars in the J-band, and 9 comparison stars in the Ks-band in each field. To estimate the photometric accuracy of this procedure, we measured the rms of brightness of a few random bright stars in the field (presumably constant stars), and obtained rms of 0.008 mag for the J-band and 0.006 mag for the Ks-band. 
For 6 photometric nights, we observed a significant number (8-12) of photometric standard stars from the United Kingdom Infrared Telescope (UKIRT) system (Hawarden et al. 2001) at a variety of air masses and spanning a broad range in colors. It allowed us to calibrate brightness on the UKIRT standard system, with the calibration uncertainty of 0.012 mag in the J-band, and 0.013 mag in the Ks-band.

\begin{figure}
    \begin{center}
        \includegraphics[width=0.49\textwidth]{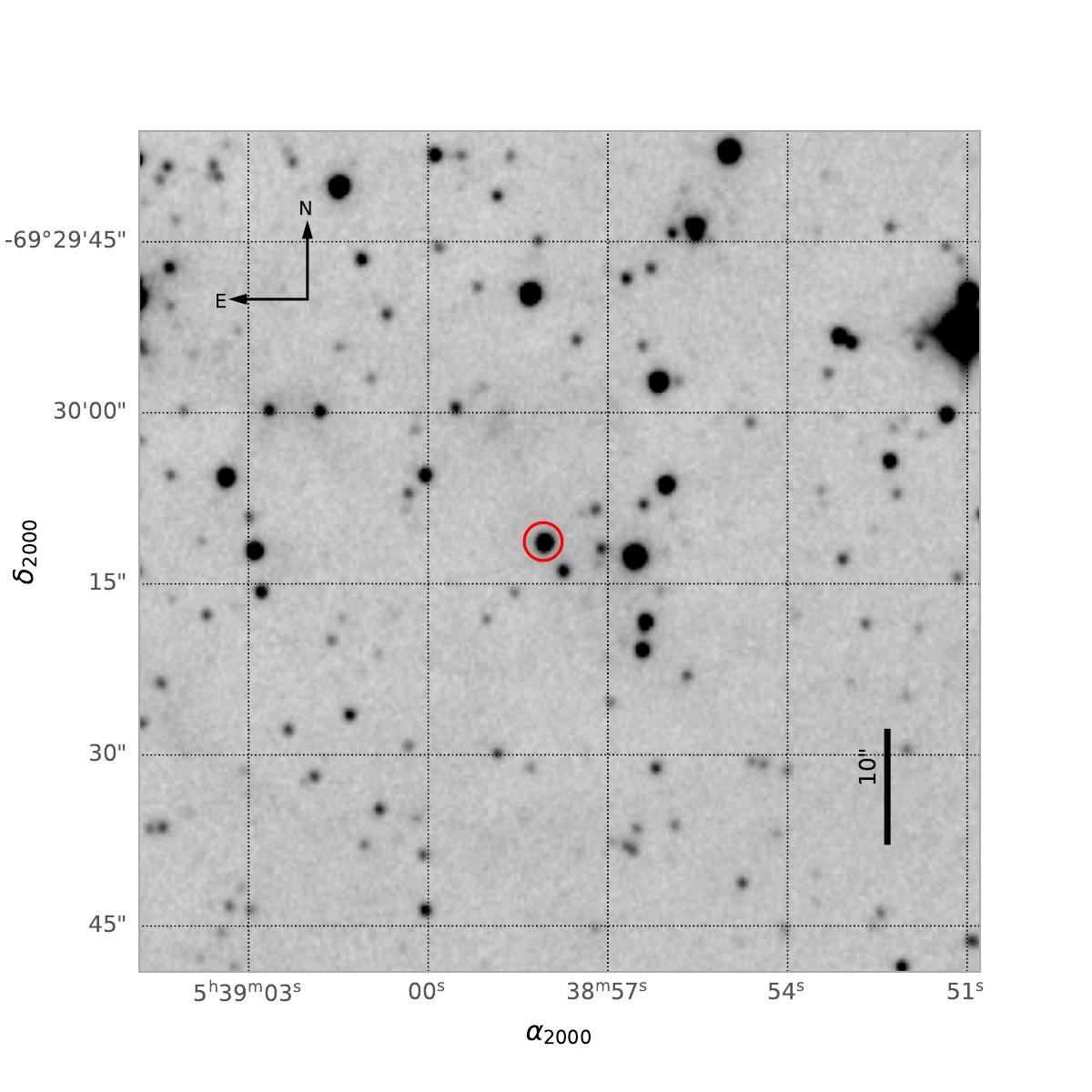}
    \end{center}
    \caption{ A part of a SOFI $K_\mathrm{S}$-band frame showing a neighborhood of the BLMC-03 system, which is marked with a red circle.}
    \label{fig:sky}
\end{figure}

In Fig.~\ref{fig:sky}, we show the neighborhood of BLMC-03 as seen on one of the SOFI Ks-band frames. The system can be easily resolved from all the other identified nearby stars.

Unfortunately, because of randomly appearing patterns in one quadrant of the camera, some of our measurements had to be discarded. However, joining the SOFI data with the VMC Ks photometry, we have both eclipses covered and the flat parts of the light curves (out of the eclipses) well-defined. As far as we know, our final light curve with 26 data points is the best-covered Ks-band light curve for an extra-galactic eclipsing binary system.

\subsection{Spectroscopy}\label{spec}

We acquired high-resolution optical echelle spectra of BLMC-03 using the UVES spectrograph \citep{dekker:2000} mounted on the Very Large Telescope (VLT) at Paranal Observatory and the MIKE spectrograph \citep{bernstein:2003} on the Magellan Clay Telescope at the Las Campanas Observatory (LCO), both in Chile. The setup of the instruments and the reduction procedure are the same as described in Paper I.
A total of 17 high-resolution spectra  (9 UVES + 8 MIKE) were obtained with a typical signal-to-noise ratio near $4000$~\r{A} of about 50-60 for UVES and 30-40 for MIKE.

To increase the time coverage of observations, we used in addition 7 archival spectra collected with the Magellan Echellette (MagE) spectrograph, also mounted on the Clay telescope at the LCO. These data were used previously by \cite{massey:2012}, where a detailed description of the instrument setup, reduction, and calibration can be found.  Thus in total 24 spectra were used for the determination of radial velocities (RVs) in a homogeneous way and in the main analysis of the binary system. 

We note that \citet{massey:2012} provide 4 more velocity determinations, which come from the analysis of their spectra acquired with the Inamori-Magellan Areal Camera and Spectrograph (IMACS) instrument on the Baade Telescope at the LCO. We took these data into account in the analysis of the outer (A+B) orbit but did not use them to obtain the final results (see section~\ref{sec:3body} for details).

\section{Analysis} \label{sec:analysis}

The analysis of BLMC-03 is similar to those performed in our previous papers (I and II) with the difference of taking into account the presence of other bodies (see Section~\ref{sec:3body} and \ref{sec:o-c} below). Similarly to the study of BLMC-02 described in Paper II we also used NLTE spectral analysis to obtain temperatures of the components and an independent estimate of their luminosity ratio.

Following the typical process for eclipsing binaries, we model the light and radial velocity curves to characterize the system and obtain the physical parameters of its components. Before the main analysis described in detail below, preliminary fits to the available data were done to obtain an approximate characterization of the system and the stars and identify possible complications. The main analysis consists of the determination of radial velocities of the primary and secondary components (free of higher-level orbital motion) and simultaneous modeling of the light and RV curves corrected for the light-travel time effect. This is followed by the NLTE spectral analysis. Finally, from the combined results, the physical parameters of the components are derived.

\subsection{Radial velocities and orbital solution}\label{sec:orbit}

We started our analysis with the determination of radial velocities of the primary and secondary components of BLMC-03. For that, we applied the Broadening Function technique \citep[BF;][]{rucinski:1992, rucinski:1999} to the reduced one-dimensional spectra using the {\tt RaveSpan} code \citep{pilecki:2017}. 
As a template spectrum, we used a spectrum of a standard star from the ESO Data Archive, that was selected to match the spectral type of the components.

Once we had extracted the radial velocities from the spectra, we modeled the RV curve taking into account the orbital elements of a spectroscopic binary: period ($P_A$), the orbital semi-amplitudes ($K_A$), the velocity of the center-of-mass ($V_A$), the reference time ($T_{0,A}$), the orbital eccentricity ($e_A$) and the argument of periastron passage of the primary ($\omega_A$). We fitted a model to the observed RV curve of each component of the system and obtained a preliminary orbital solution.

\subsubsection{Third body}\label{sec:3body}

In the preliminary RV curve analysis, we detected systematic and correlated changes in the residual RVs for the primary and secondary components with a period of about 200 days. This was a clear indication of the inner (Aa+Ab) system orbiting the common center of mass (COM) of the higher-order system. As from the light curve analysis, we also obtained significant values for the third light\footnote{The parameter known as a "third light" in binary modeling describes any additional light that does not come from the components of the eclipsing binary system.} in all considered bands, we concluded that the system is at least a triple, and included the presence of a third body in the RV analysis. In the final model of the Aa+Ab binary, we thus included additional parameters that describe the orbital motion of this binary around the A+B system COM. We then fitted this model to the measured velocities of the primary and secondary components.

\begin{figure}
        \includegraphics[width=0.48\textwidth]{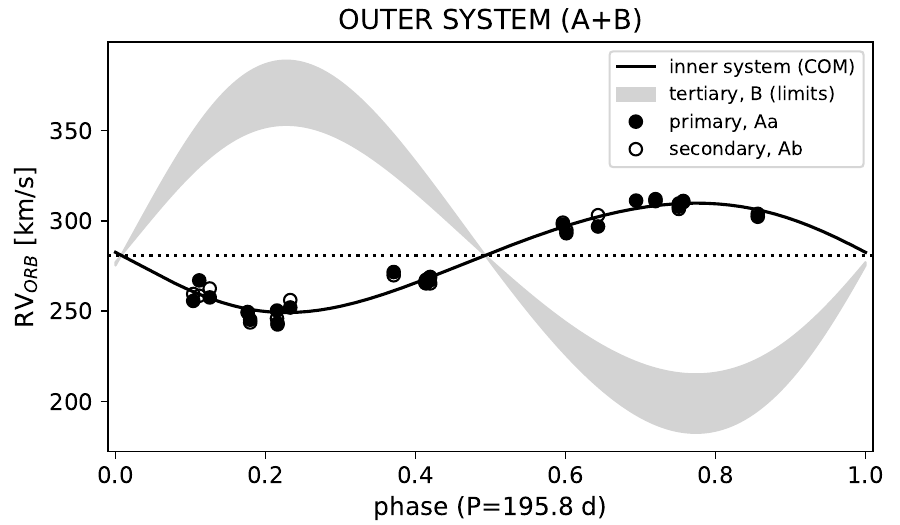}
    \caption{Radial velocity of the inner (Aa-Ab) system center-of-mass and the range of possible velocities of the tertiary component (assuming it is single for the superior limit). Velocities of the primary and secondary components are residuals after the subtraction of RVs from the inner binary model.}    
    \label{fig:outer}
\end{figure} 

To increase the number of data points, we considered adding to the analysis radial velocities measured by \cite{massey:2012} from their spectra obtained with IMACS as mentioned in Section~\ref{spec}. These radial velocities were measured using a low number of spectral lines: 1, 3, 4, or 6 depending on the spectrum \citep[see table 6 of][]{massey:2012}.
We considered the measurement using just one line as not reliable and indeed a quick check showed the difference between the primary and secondary velocities to be about 50 km~s$^{-1}$ too low. The uncertainties for them are also not provided. We thus used in the analysis only measurements based on at least 3 lines, with determined uncertainties.
After doing the fit, we noted however that these literature RVs, although being marginally consistent within error bars, are all lower than the best model by about 15-20 km~s$^{-1}$. Moreover, this shift is also seen when comparing one of these points with our two measurements done for spectra taken 5 days later. As the data we used for our RV measurements dominate in number and quality the exclusion of these external measurements leads to practically the same results regarding the A+B system (amplitude change of 0.3$\sigma$), and more importantly, negligible changes in the parameters of the inner one (below 0.2$\sigma$). Taking all of this into account and to maintain the homogeneity of measurements we decided to use, here and in the rest of the analysis, only RV measurements determined by us (using uniformly the BF method) from the UVES, MIKE, and MagE spectra.
The final orbital solution for the outer system is shown in Fig.~\ref{fig:outer} and the properties of its orbit are given in Table~\ref{tab:3rdorb}. 

\begin{deluxetable}{lcc}
    \tabletypesize{\footnotesize}
    \tablecaption{Orbital elements of the outer A+B orbit of BLMC-03}\label{tab:3rdorb}
    \tablehead{\colhead{parameter} & \colhead{unit} & \colhead{value}}
           \startdata
            $P_{B}$        & [days]  &  195.78 $\pm$ 0.13  \\
            $T_{0,B}$      & [days]  &  8038.3 $\pm$ 1.0   \\
            $e_{B}$        &      &   0.086 $\pm$ 0.024 \\
            $\omega_{B}$   & [deg]  &    123 $\pm$ 15     \\
            $K_{A}$        & [km s$^{-1}$] & 30 $\pm$ 0.6 \\
            $K_{B,min}$     & [km s$^{-1}$] &  69$^*$ \\
            $K_{B,max}$     & [km s$^{-1}$] &  103 \\
           \enddata    
\tablenotetext{*}{Assuming the third object is a single star.}
\end{deluxetable}

The inclusion of the third body in the orbital solution has decreased the scatter (rms) of residuals from about 22 to 4 km~s$^{-1}$.  The orbital period of the outer system converged firmly at $P_{A+B} = 195.78 \pm 0.13$ days, resulting in the period ratio between the outer and inner system of $P_{A+B}/P_{A}\approx 61$. A mass estimate for the tertiary will be derived later, once we have the final masses of the inner components. The main result from this part of the analysis is radial velocities of the primary and secondary components free of additional orbital movement around the outer A+B system COM and a preliminary orbital solution, which will be used as an input in the final modeling.

\subsubsection{Stellar rotation}\label{sec:rota}

\begin{figure}
    \includegraphics[width=0.48\textwidth]{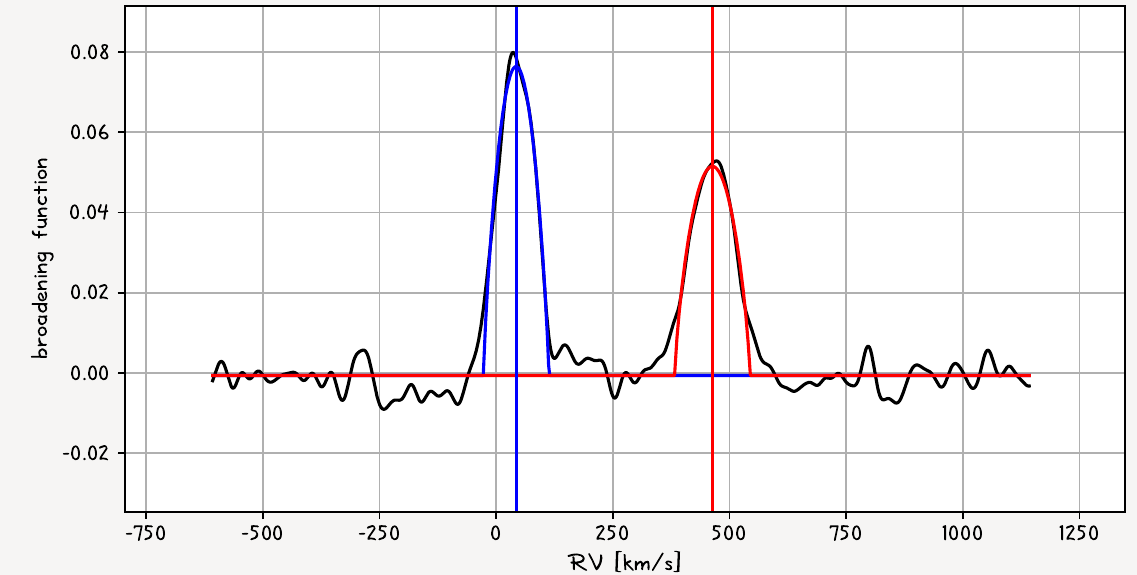} 
\caption{An example of Broadening Function profile for one of the analyzed spectra. Apart from radial velocities, rotational velocities are measured, which are lower than typical for early-type stars.}
\label{rotation}
\end{figure}

From the BF method, the projected rotation velocities of the components of the system ($v_{1,2} \sin i$) were also obtained. The broadening function response for one of the spectra is presented in Fig~\ref{rotation}.
We have determined $v_{1,2} \sin i$ for various spectra taken close to the orbital quadratures, where the profiles are well separated, and taken an average. To obtain the true rotation velocities we assume the inclination of the rotation axes to be the same as the orbital inclination and correct measured velocities for the rotation velocity of the template star. We did not correct for instrumental broadening because it is negligible compared with the measured velocities.
The rotation velocities, obtained as described, are about 70 km~s$^{-1}$ and 80 km~s$^{-1}$ for the primary and secondary components, respectively.

The final values are later used in the modeling in the {\tt Phoebe} code \citep{prsa:2005} as a synchronization parameter ($F$), defined as the ratio between the rotational and orbital angular velocity. This information is important for the modeling, as rotation changes the shape of stars, which influences the light (mostly) and radial velocity (to a lower extent) curves.
For both components measured rotational velocities are sub-synchronous ($F_1$=0.61, $F_2$=0.75). This means that either their rotation is slower than the angular orbital motion or their rotation axes have inclinations lower than the orbital one. Note that we do not compare our rotation velocities with a pseudo-synchronous velocity (defined by the orbital angular velocity at periastron) as the difference between the synchronous and pseudo-synchronous velocity for very low eccentricity of \mbox{BLMC-03} is insignificant.

\subsection{Light and RV curve modeling}\label{sec:finalmodel}

The rest of the analysis closely resembles the one for BLMC-02 described in Paper I and II, where a detailed description is provided. In short, we have used a {\tt Python} wrapper to the {\tt Phoebe 1} code to obtain a simultaneous solution to the observed light and radial velocity curves. For that, we used a Markov chain Monte Carlo analysis which provides also reliable uncertainties and lets us find correlations between fitted parameters. The initial parameter values were taken from the aforementioned preliminary low-resolution analysis. A slight difference from the standard analysis was the usage of radial velocities of the components corrected for the orbital motion around the outer system's center of mass (in comparison to using the velocities directly). 

\begin{figure}
        \includegraphics[width=0.45\textwidth]{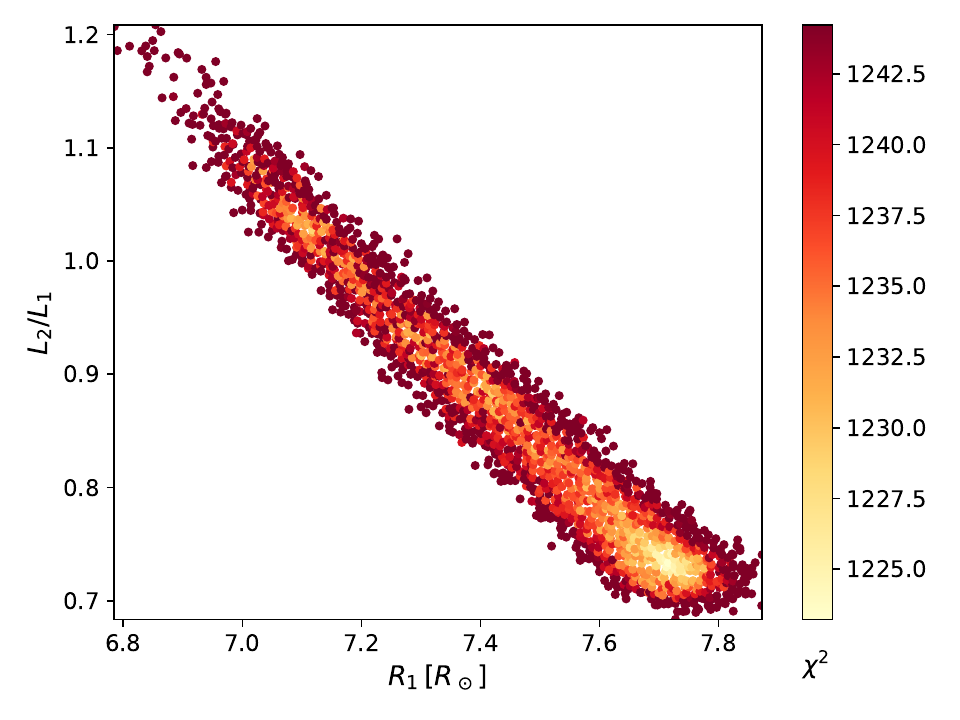}
        \includegraphics[width=0.45\textwidth]{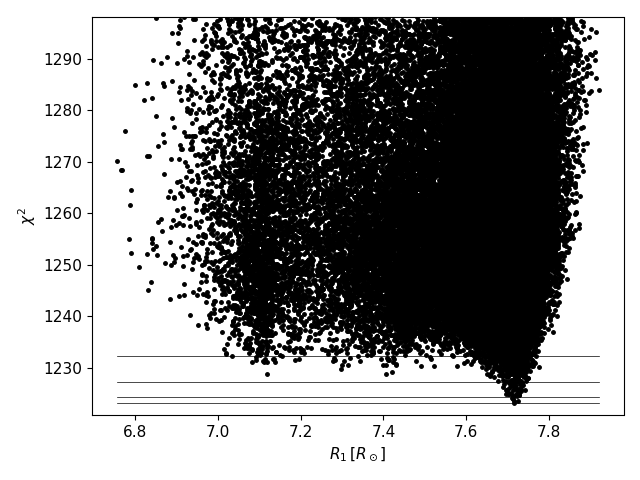}
    \caption{Top: Results from the Markov chain Monte Carlo analysis showing a correlation between the radius of the primary and the luminosity ratio. Bottom: The $\chi^2$ values vs. radius of the primary for the models calculated in the analysis.}
    \label{fig:degeneracy}
\end{figure}

As typical for well-detached, non-totally eclipsing systems we found a slight degeneracy between the parameters of the component stars. This can be seen, for example, in the correlation between the radius of the primary ($R_1$) and the luminosity ratio ($L_2/L_1$) in Fig.~\ref{fig:degeneracy} (top). The solution is well-defined (with small, symmetric uncertainties) within 1$\sigma$ but strongly asymmetric for 3-$\sigma$ uncertainties as seen in the $\chi^2$ plot for $R_1$ in the bottom panel. However, we note that along the $L_{2}/L_{1}$ - $R_1$ relationship (top panel) the flux ratio $F_{2}/F_{1}$ = $(R_{1}/R_{2})^2(L_{2}/L_{1})$ remains constant. We measure $F_{2}/F_{1}$ = 0.971$\pm$0.002 in the V-band and 0.968$\pm$0.004 in both I$_{c}$ and K$_{s}$ bands. This flux ratio will be used to constrain the effective temperature difference between the primary and secondary components in the non-LTE spectral analysis described in Section~\ref{sec:spectral}, which in turn will be used to obtain the final, non-degenerate solution.

\begin{figure}
    \begin{center}
        \includegraphics[width=0.48\textwidth]{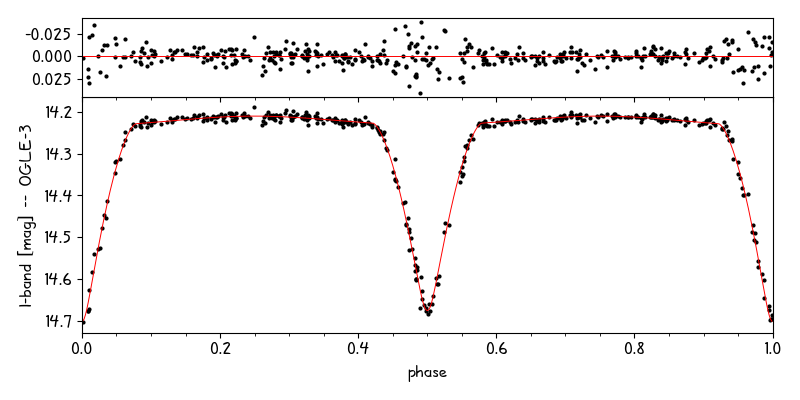} 
    \end{center}
    \caption{$I_\mathrm{C}$-band light curve for BLMC-03 not taking into account the presence of the 4th component. The best model fit is shown as a red line. A high residual scatter can be seen around eclipses.}
    \label{fig:lcs_scat}
\end{figure}

\subsection{O-C analysis} \label{sec:o-c}

\begin{figure}
    \begin{center}
      \includegraphics[width=0.49\textwidth]{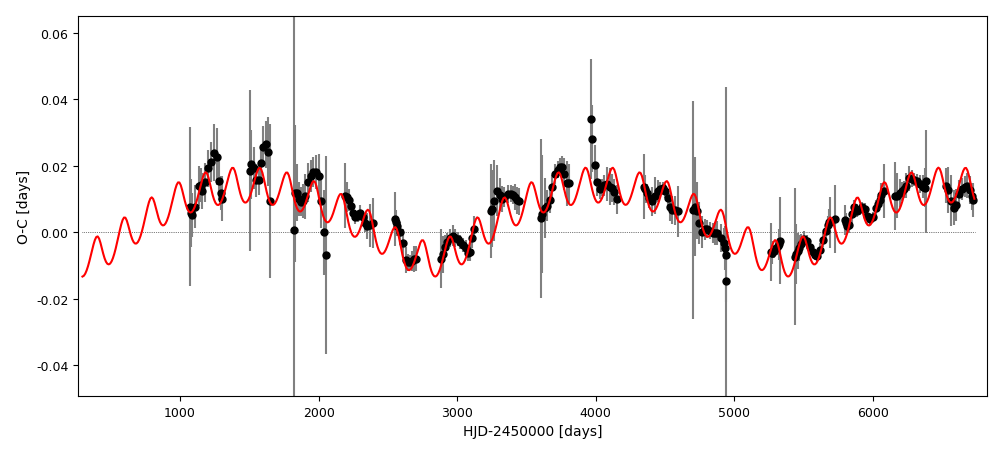}
    \end{center}
    \caption{O-C diagram showing the phase shifts between the data from around the given date and the average model. The influence of two extra components can be seen. The red line is the fit to the LTTE induced by the 3rd and 4th body. }
    \label{fig:oc}
\end{figure}

Looking at the light curve models obtained in the previous section we found a significant residual scatter around eclipses (see Fig.~\ref{fig:lcs_scat}).
To find out the cause of this scatter and trying to independently check the orbital parameters of the third body (B), we performed the O-C analysis similar to that described in Pilecki et al. (2021) using the long time-span light curves from the EROS-2 and OGLE projects. 
The strongest variation seen in the O-C diagram (see Fig.~\ref{fig:oc}) indeed looked like caused by the orbital motion but to our surprise, its period was determined as 2550 $\pm$ 30 days. The shorter-period variability previously detected in RVs was also found with a period of $195.3 \pm 0.5$ days but with much lower amplitude, consistent with the spectroscopic orbit determined in Section~\ref{sec:3body} ($K_{A,O-C} = 26 \pm 3$ compared to $K_{A,RV} = 30.0 \pm 0.6$). This variability is best seen in the dataset of OGLE-IV (HJD $>2455000$ days) which has the highest density and quality of measurements.
Both variabilities can be described by the light-travel time effect (LTTE), as is shown by the red line fit in Fig.~\ref{fig:oc}. Apart from that, no other secular, cyclic, or erratic changes were detected.

\begin{figure*}
    \begin{center}
        \includegraphics[width=0.49\textwidth]{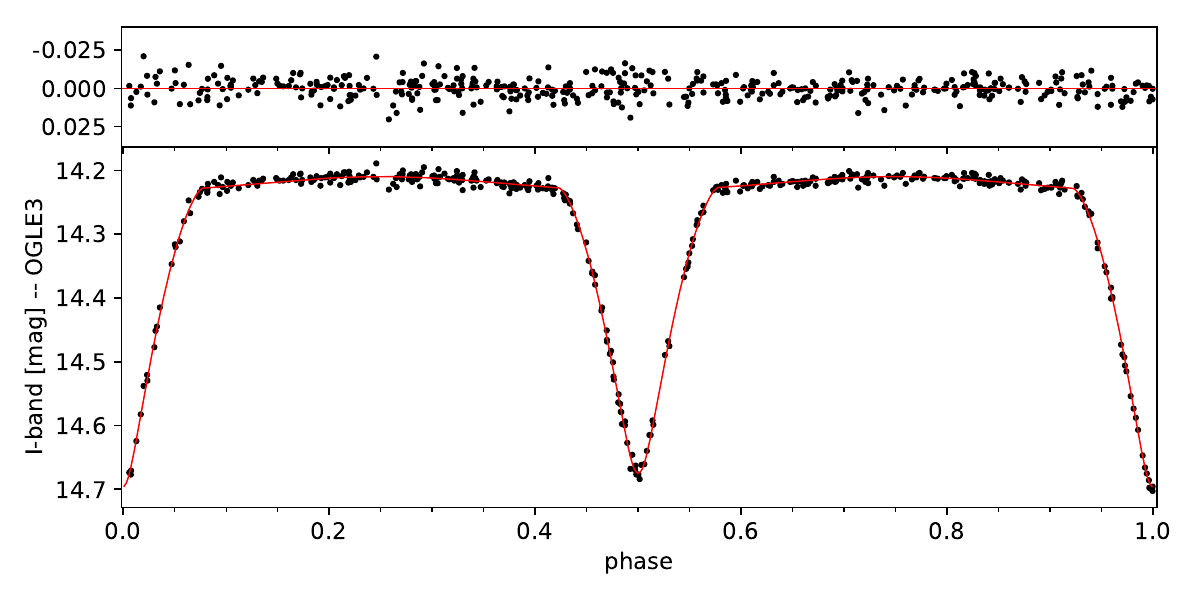} 
        \includegraphics[width=0.49\textwidth]{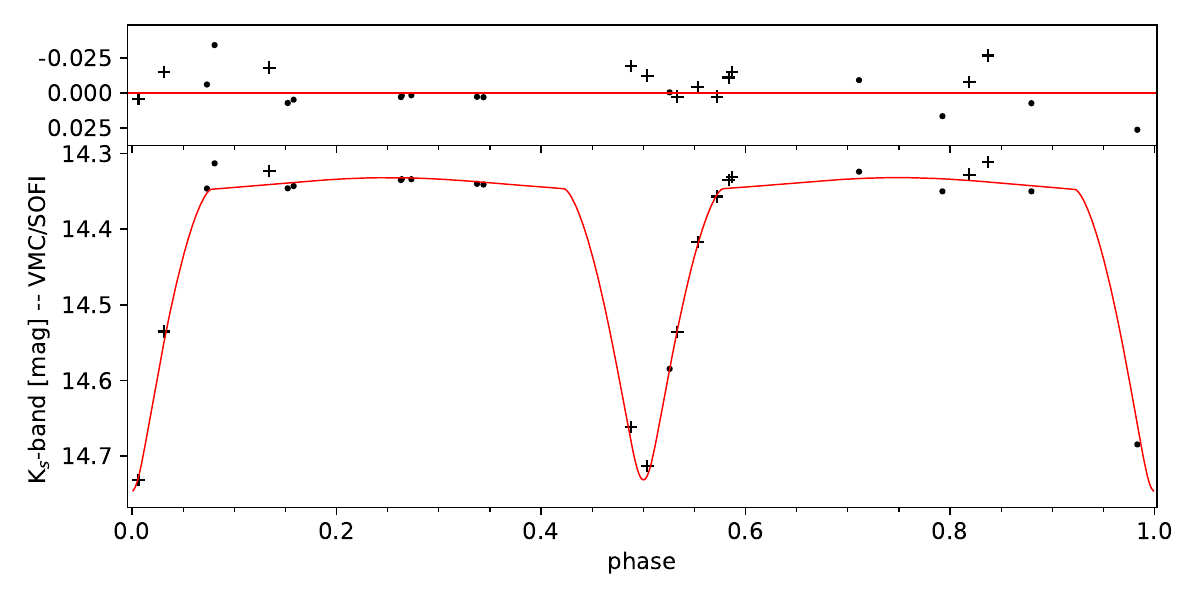}
    \end{center}
    \caption{$I_\mathrm{C}$- and $K_\mathrm{S}$-band light curves for BLMC-03 corrected for the gravitational influence of the 4th component. The residual scatter around eclipses disappeared almost completely. In the right panel, VMC and SOFI data are represented by points and crosses, respectively.}
    \label{fig:lcs}
\end{figure*}

\begin{figure}
    \begin{center}
        \includegraphics[width=0.48\textwidth]{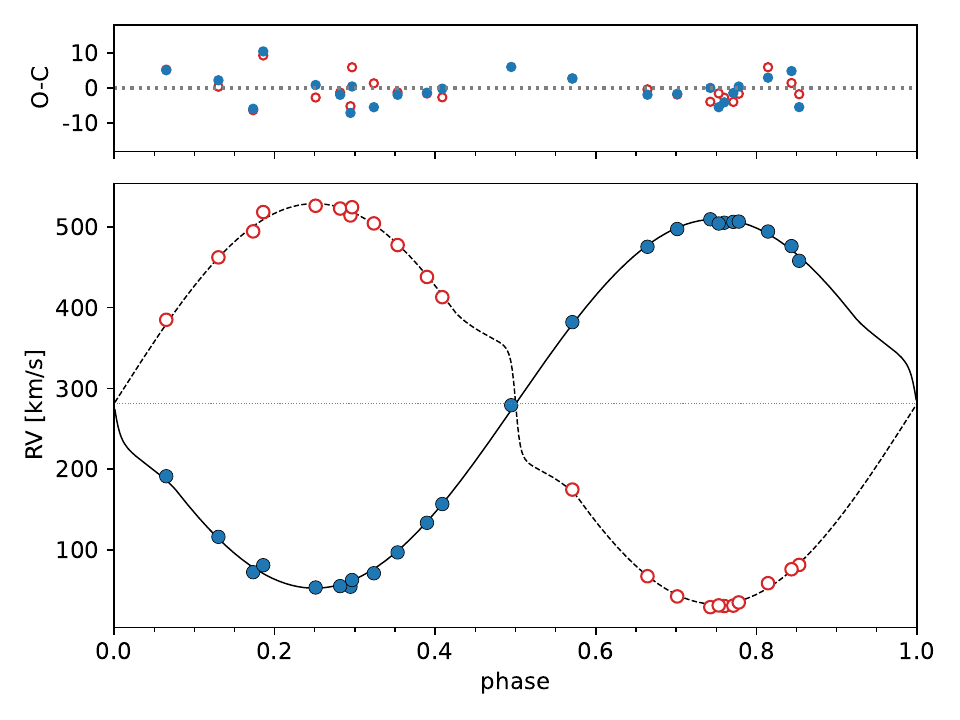}
    \end{center}
    \caption{Orbital solution for the inner binary system. Measured radial velocities for the primary (blue) and secondary (red) are shown with residuals in the upper panel.}    
    \label{fig:rvc}
\end{figure}

With these results, we could not only confirm the existence and the orbital period of the third component found from the analysis of radial velocities but also the presence of a fourth component. The system is thus at least a quadruple with the configuration (Aa+Ab)+B+C. Because of a wider orbit, the influence of the 4th body (C) on the phase shift of the eclipsing variability of the A system is much stronger than that of the 3rd body (B). To account for this periodic phase shift we subtracted it from the light and radial velocity curves by adjusting the measurement time by the value obtained from the LTTE fit. We then modeled these corrected data and obtained new model fits. As can be seen in Fig.~\ref{fig:lcs}, the extra scatter in the eclipses disappeared almost completely. As the depths and shape of the eclipses are now better defined, some parameters, like the radius, temperature ratio, and the third light parameter changed slightly. The final fitted values are close to the ones determined before (within uncertainties) but are more precise. The degeneracy also decreased. The final radial velocity curve is shown in Fig.~\ref{fig:rvc}.

\subsection{Spectral analysis and reddening determination} \label{sec:spectral}

The purpose of the non-LTE spectral analysis of the observed binary is primarily to determine the effective temperatures of the components and interstellar reddening and extinction. This will enable us to discuss the physical nature and evolutionary status of the components and to constrain interstellar reddening and extinction, providing intrinsic colors and surface brightnesses as well as an independent estimate of the distance.
We also used this analysis to independently find the best solution along the slight degeneracy described in Section~\ref{sec:finalmodel}.

We proceed in a very similar way as in Paper II. We use the non-LTE model atmosphere code {\tt FASTWIND }\citep{Puls2005, Rivero2012} and calculate \textit{normalized} flux spectra F$_{1,\lambda}$ and F$_{2,\lambda}$ for the primary and secondary star, respectively, and then combine those to a composite spectrum through

\begin{equation}
  F_{\lambda} = w_1F_{1,\lambda}(T_1, log~g_1, v_1) + w_2F_{2,\lambda}(T_2, log~g_2, v_2) + w_3.
\end{equation}

\noindent $T_i$, $log \, g_i$ and $v_i$ denote effective temperature, gravity, and radial velocity of the primary and secondary. The weight $w_3$ accounts for the contribution of the additional light (the third light parameter in the binary model) described in the previous sections and is defined as

\begin{equation}
w_3 = {L_3 \over L_1 + L_2 + L_3},
\end{equation}

\noindent where $L_1$ and $L_2$ are the V-band luminosities of the primary and secondary, and $L_3$ is the total luminosity of all additional components. From the photometry fits discussed above we obtain $w_3 = 0.19 \pm 0.01$. The weights $w_1$ and $w_2$ represent the continuum contributions by the primary and secondary, respectively, and are given by

\begin{equation}
w_1 = {L_1 \over L_1 + L_2 + L_3} = {1 - w_3 \over 1+ (R_{2}/R_{1})^2(F_{2}/F_{1})}
\end{equation}

\noindent and

\begin{equation}
w_2 = {L_2 \over L_1 + L_2 + L_3} = 1 - w_3 - w_1.
\end{equation}

We use the observed flux ratio given in  section~\ref{sec:finalmodel} for the calculation of $w_1$ and $w_2$, together with the primary and secondary radii and stellar gravities along the solutions seen in Fig.~\ref{fig:degeneracy} as obtained from the light curve and radial velocity curve modeling described in previous sections. 

\begin{figure}
    \begin{center}
      \includegraphics[width=0.45\textwidth]{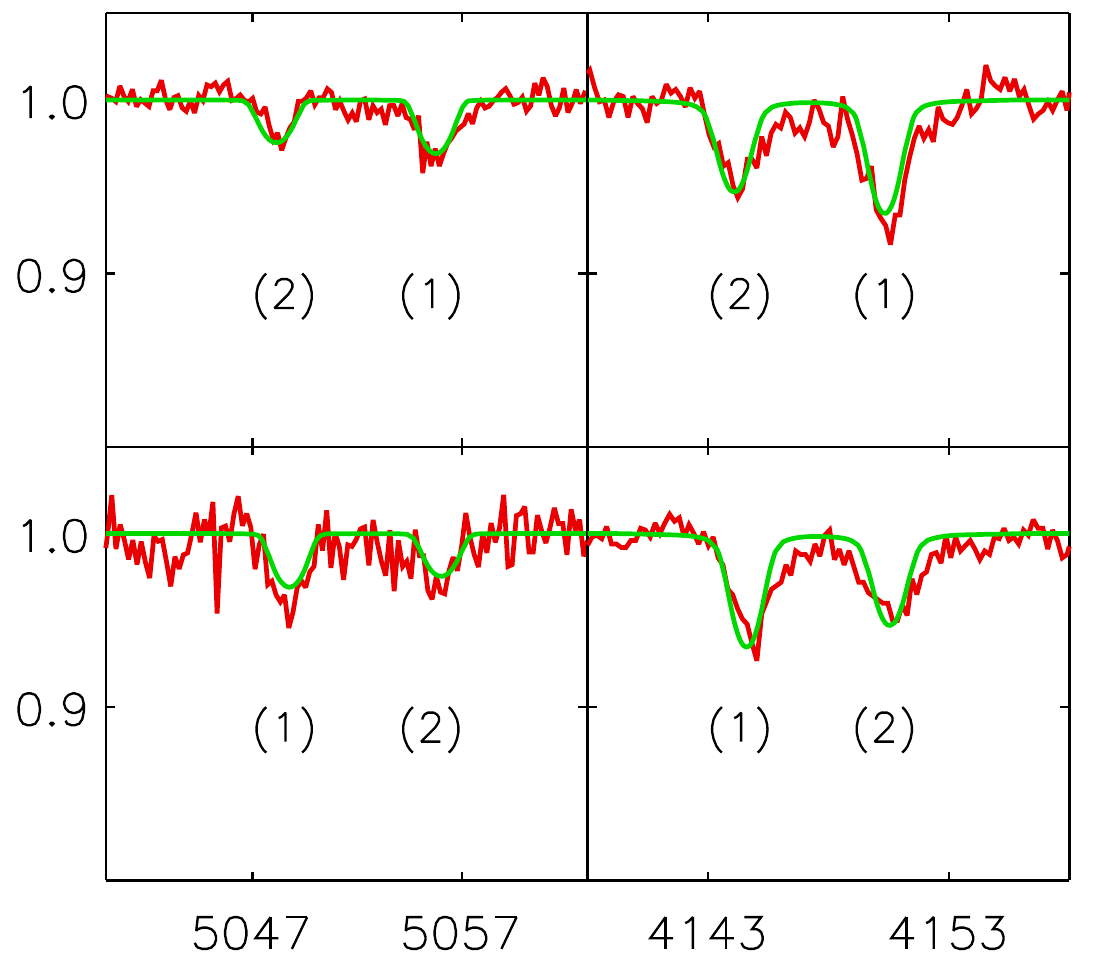}
      \includegraphics[width=0.45\textwidth]{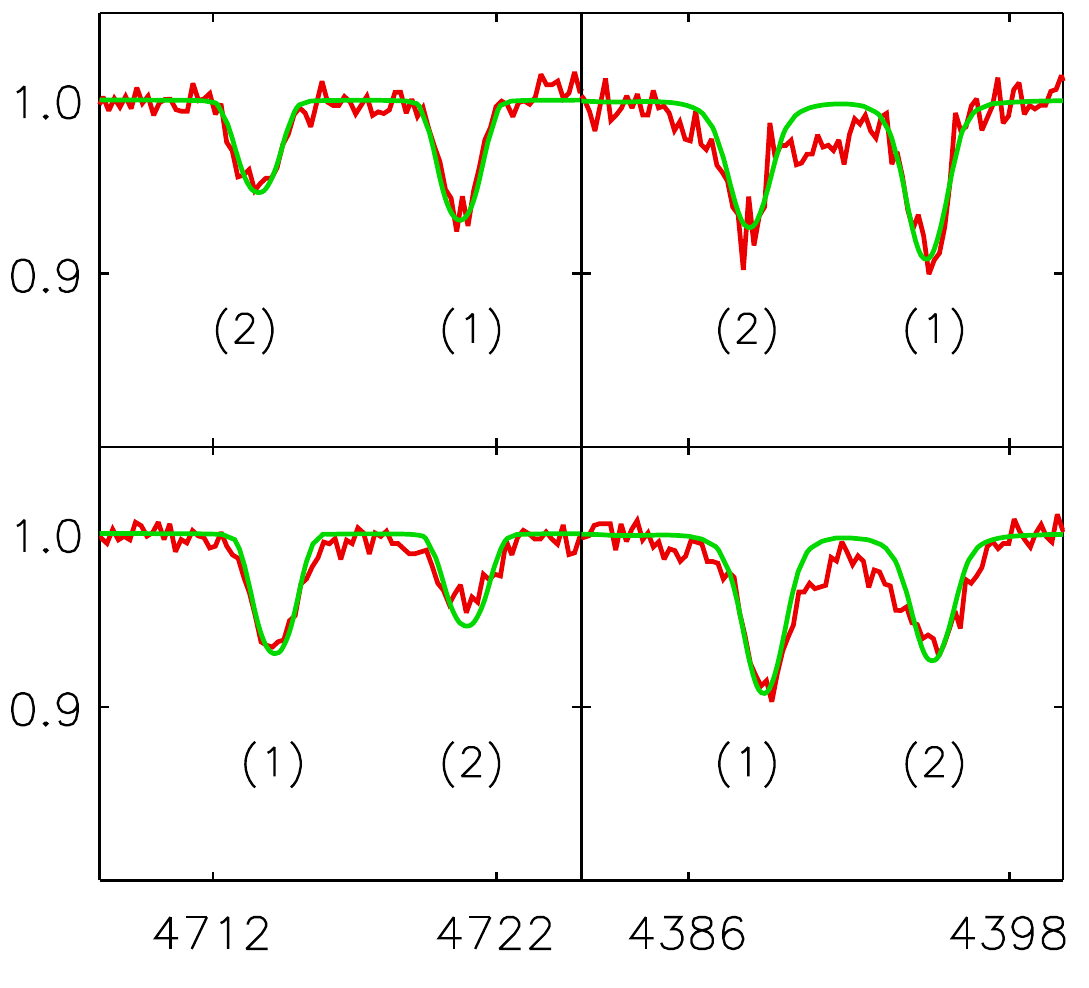}
    \end{center}
    \caption{Fit of HeI lines with the best composite binary model spectrum obtained from the $\chi^2$-analysis. The upper figure shows HeI 5048 (left) and HeI 4143 (right) and the lower figure HeI 4713 (left) and HeI 4387 (right). The upper row in each figure corresponds to the superposition of three UVES spectra around phase 0.75 and the lower row to three MIKE spectra at phase 0.2. The line contributions from the primary and secondary are indicated by (1) and (2), respectively.}
    \label{fig:HeI}
\end{figure} 

\begin{figure}
    \begin{center}
      \includegraphics[width=0.45\textwidth]{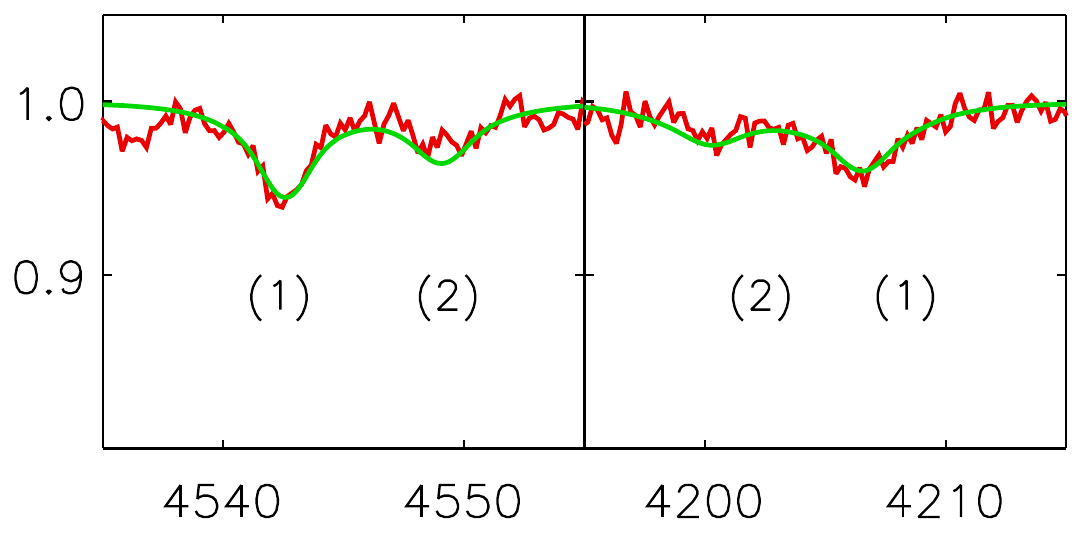}
      \includegraphics[width=0.45\textwidth]{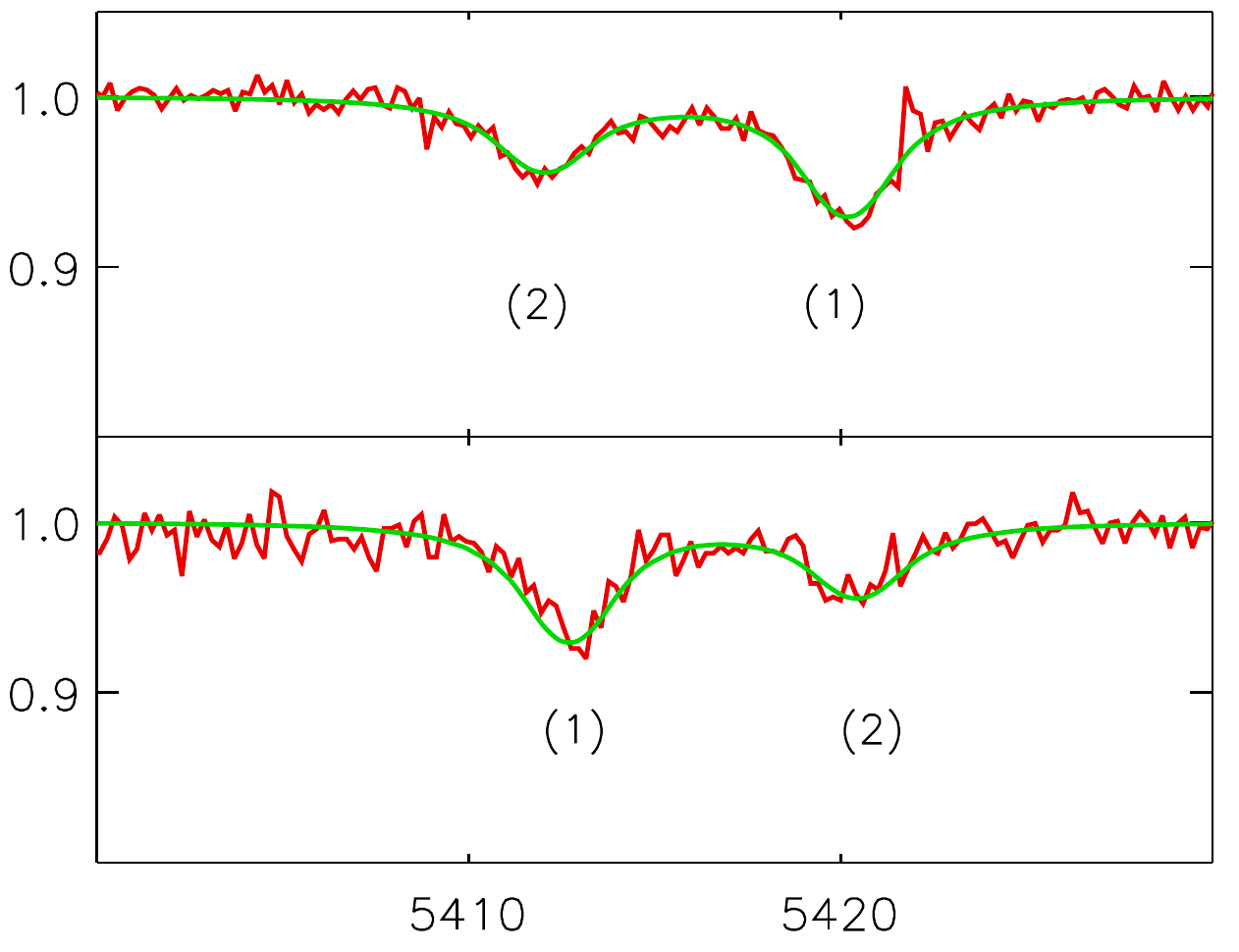}
    \end{center}
    \caption{Fit of HeII lines. The upper figure shows the MIKE spectrum of HeII 4541 (left) and the UVES spectrum of HeII 4200. The lower spectrum displays fits of the UVES (upper row) and MIKE (lower row) spectra of HeII 5411.}
    \label{fig:HeII}
\end{figure} 

In the first step, we determine the effective temperature difference $\Delta T = T_1 - T_2$ between primary and secondary. We use {\tt FASTWIND} model SEDs to calculate stellar V-band fluxes and compare their ratio with the observed value. For the gravities from the binary model (as mentioned above) and with temperatures between 30000 to 39000K the observed ratio leads to $\Delta T = 500\pm100$ K. We then calculate a grid of composite spectra close to binary quadrature with this temperature difference $\Delta T$ varying the primary temperature between 31000 to 39000 K for each pair of radii and gravities. As in Paper II, we adopt four values of the helium abundance N(He)/N(H), 0.08, 0.09, 0.1, and 0.15, two values for the microturbulence v$_{turb}$ = 10 and 15 km~s$^{-1}$ and metallicity of [Z] = log Z/Z$_{\odot}$ = -0.35 \citep{urbaneja:2017}. We calculate normalized spectra with line profiles accounting for the observed rotational velocities and the spectral resolution of the spectrographs by convolving with the corresponding broadening profiles.

\begin{figure*}
  \begin{center}
    \includegraphics[width=0.4\textwidth]{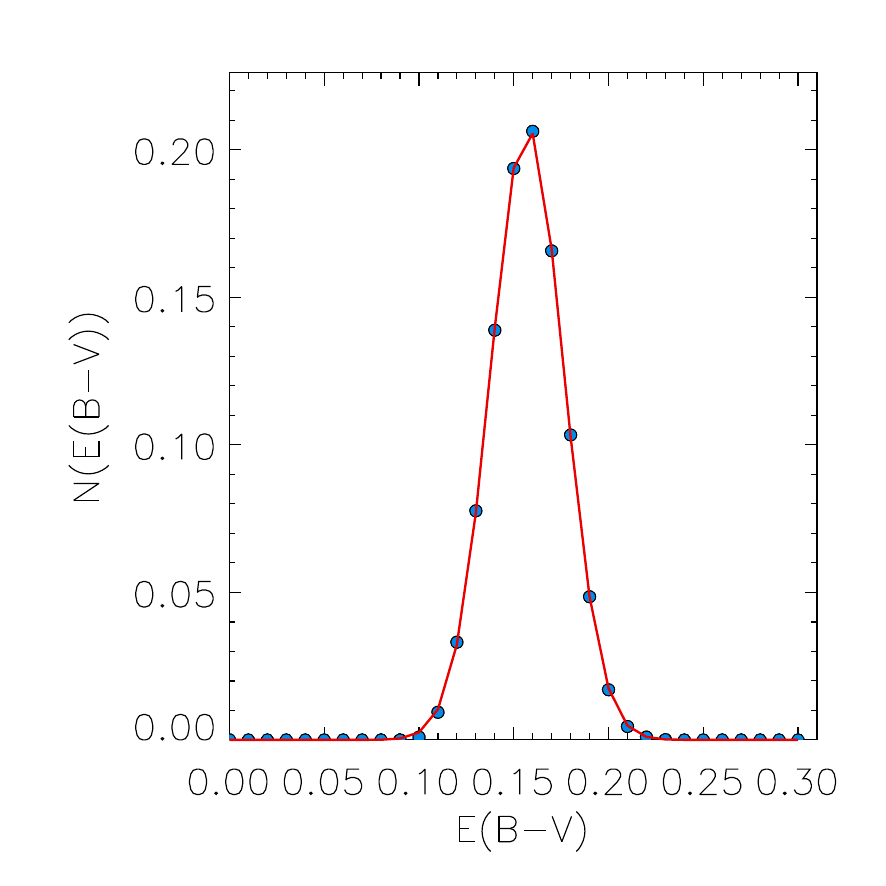}
    \includegraphics[width=0.4\textwidth]{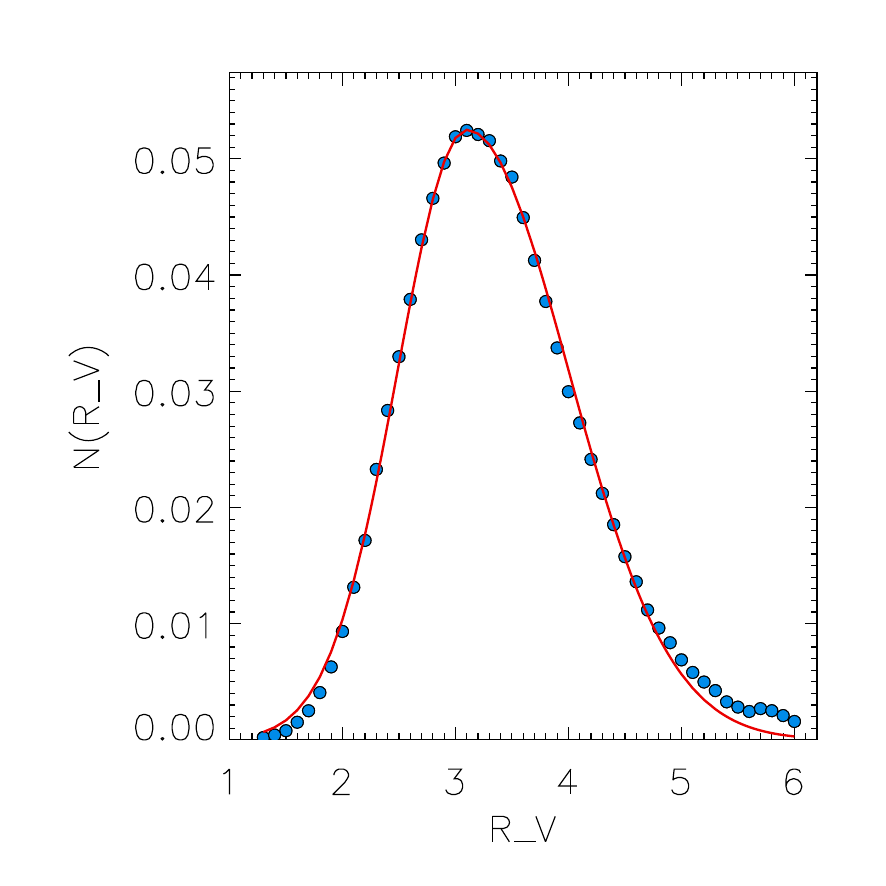}
    \includegraphics[width=0.4\textwidth]{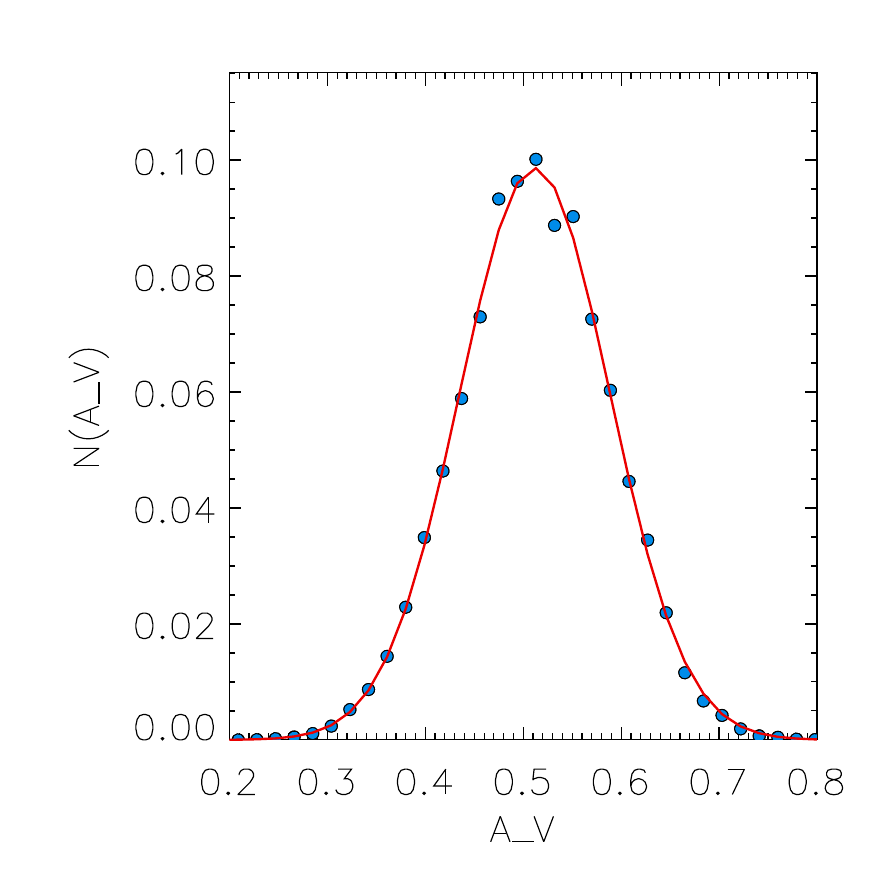}
    \includegraphics[width=0.4\textwidth]{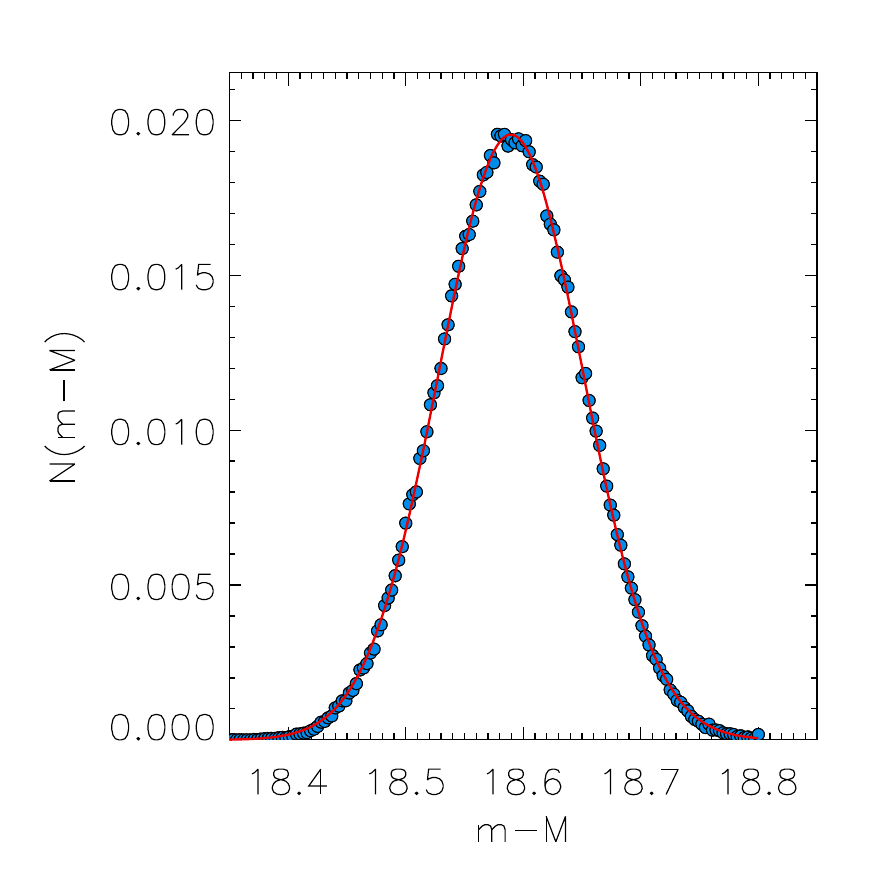}
  \end{center}
  \caption{Probability distributions (blue circles) and Gaussian fits (red) obtained from the Monte Carlo fit of observed colors as described in the text. The red curves are Gaussian fits. Top left: N(E(B-V)), top right: N(R$_V$), bottom left: N(A$_V$), bottom right: N(m-M).} \label{fig:probdm}
\end{figure*}

In our analysis, we proceed exactly as in paper II. We select the same
set of HeI and HeII lines and compare model and observed spectra by
calculating $\chi^2$-values for all radii, temperatures, helium
abundances, and microturbulence values and use the minimum of $\chi^2$
combined with Monte Carlo simulations to constrain the stellar
parameters of the system. We obtain
$34250\pm500$ and $33750\pm500$ K for the
effective temperatures of the primary and secondary, respectively, together with a helium abundance
N(He)/N(H)=0.09$\pm0.01$ and a microturbulence v$_{turb}$ = 10 km~s$^{-1}$.
Our best solution is obtained for the primary radius of $R_1 = 7.7 R_{\odot}$ which confirms the best solution from the binary modeling. Examples of the fits of the helium lines with the final model are shown in Fig.~\ref{fig:HeI} and Fig.~\ref{fig:HeII}.

\subsubsection{Reddening and extinction}

In the next step, we determine the interstellar reddening and extinction. This is crucial for an independent estimate of distance and a measurement of intrinsic stellar surface brightness and colors. For this purpose, we use multiband photometry outside the eclipses, corrected for the contribution of the additional components. In the Johnson-Cousins system, we have the literature apparent magnitude B = 14.17$\pm$0.05 mag \citep{massey:2002b,zaritsky:2004,testor:1998}, as well as V = 14.34$\pm$0.03 mag and I$_c$ = 14.46$\pm$0.03 mag from our light curve modeling (the system brightness at phase 0.15).
The applied corrections for the additional light were 0.18,
  0.20, and 0.24 mag, respectively.  Using our binary model we also derived consistently out-of-eclipse \emph{Gaia} magnitudes from available light curves \citep[DR3,][]{gaia3:2022} with   G$_B$ = 14.22 $\pm$ 0.03 mag and G$_R$ = 14.38 $\pm$ 0.030 mag, where 3rd light corrections of  0.20 and 0.23 mag were applied. From our modeling, we also derived the near-IR magnitudes of BLMC-03 in J (VMC data) and K$_s$-band (VMC+SOFI). We adopt J = 14.61$\pm$0.04 mag and K$_s$ = 14.70$\pm$0.05 mag with 3rd light corrections of 0.28 and 0.36 mag, respectively.

We then proceed exactly in the same way as in Paper II and use the composite SED of our final model of primary and secondary to calculate reddened model colors (B-V), (G$_B$-V), (V-G$_R$), (V-I$_C$), (V-J), (V-K$_s$) in the same passbands as observed assuming a grid of E(B-V) and R$_V$ = A$_V$/E(B-V) values in
a range from 0.0 to 0.3 mag and 1.0 to 7.0, respectively. From the $\chi^2$ comparison between calculated and observed colors, we determine the best pair of E(B-V) and R$_V$ at the minimum of
$\chi^2$.  In a Monte Carlo simulation, we modify the observed colors assuming a Gaussian distribution of uncertainties and construct
probability distribution functions N(E(B-V)), N(R$_V$), N(A$_V$), and
N(m-M) for reddening, the ratio of extinction to reddening,
extinction, and distance modulus, respectively. Fig.~\ref{fig:probdm}
shows the result assuming the monochromatic reddening law by
\cite{Maiz2014,Maiz2017}. We have also applied the reddening laws by
\cite{Cardelli1989}, \cite{Odonnell1994}, and \cite{Fitzpatrick1999}
which yield very similar results. We obtain as average values E(B-V) = 0.16 $\pm$ 0.02 mag, R$_V$ = 3.2 $\pm$ 0.8,  A$_V$ = 0.51 $\pm$ 0.08 mag and a distance modulus m-M = 18.59 $\pm$ 0.06 mag. 

\begin{figure}
    \begin{center}
      \includegraphics[width=0.45\textwidth]{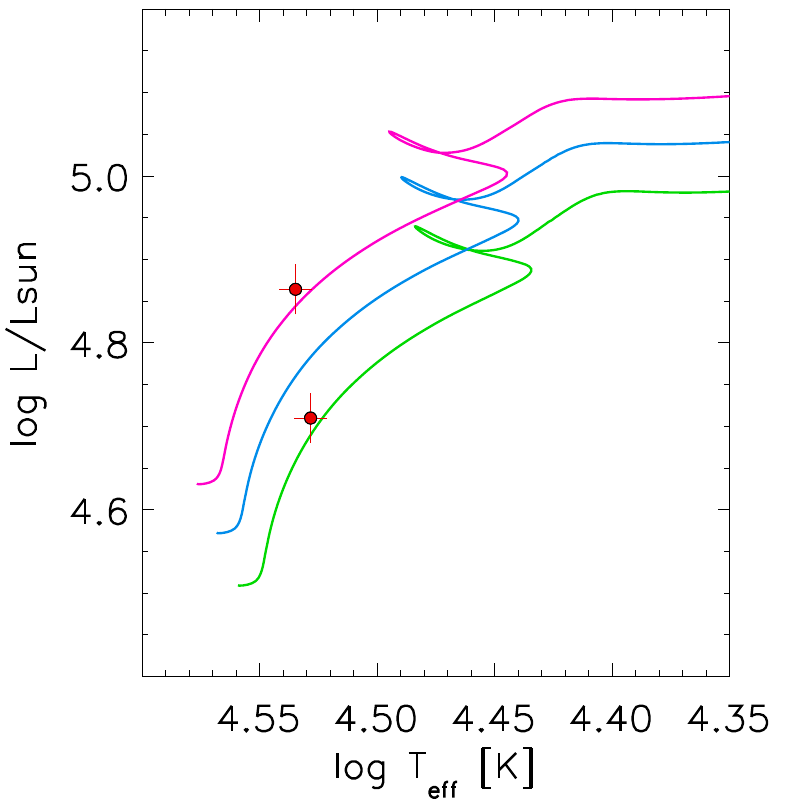}
    \end{center}
    \caption{Location of the binary components in the HRD compared to MESA evolutionary tracks with ZAMS masses of 18 (green), 19 (blue), and 20 (pink) M$_{\odot}$. }
    \label{fig:HRD}
\end{figure}

\subsubsection{Reddening from the sodium line}

\begin{figure*}
    \begin{center}
      \includegraphics[width=0.95\textwidth]{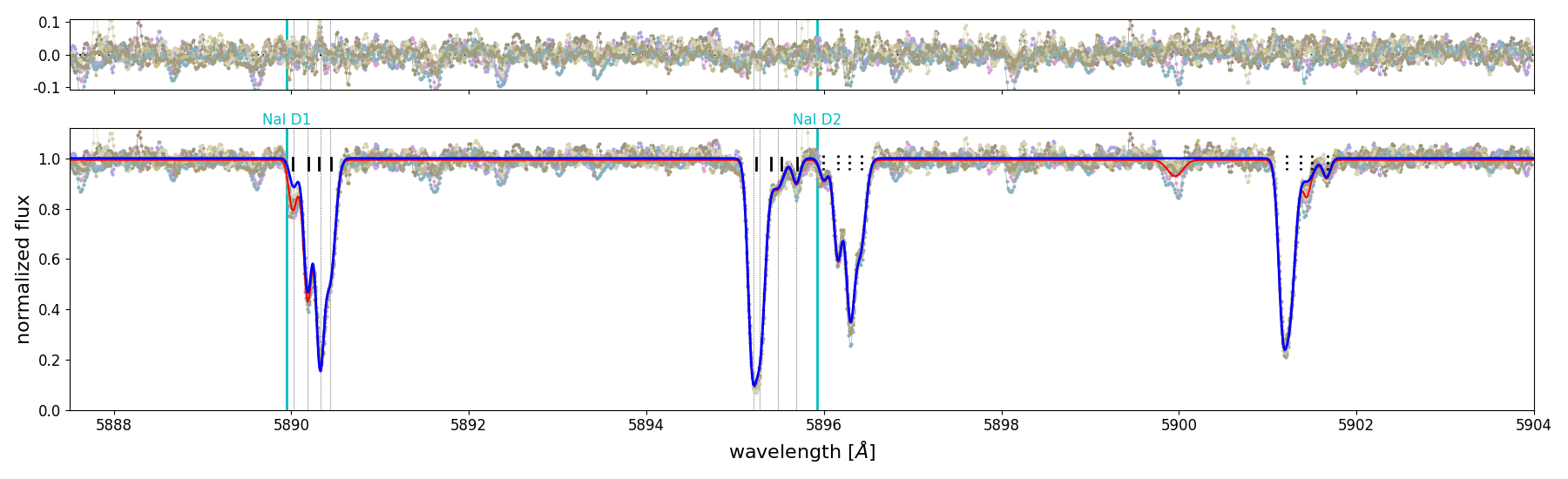}
     \end{center}
    \caption{A multi-component fit to the sodium doublet at 5890\AA. Individual components are marked with black thick vertical lines. For both the LMC and MW, four components were identified. The red line shows the fit to the data including extra lines, while the blue one marks the profiles of the sodium lines only. Multiple spectra (lines of different colors) are fitted at once. The resulting reddening E(B-V) is \NaEbv.}
    \label{fig:sodiumline}
\end{figure*}

Similar to the systems analyzed in previous papers, we did an independent determination of the reddening value through the measurement of the equivalent widths of the sodium doublet lines. The fit is shown in Fig.~\ref{fig:sodiumline}. From the calibration of
\citet{munari:1997} we then determined the reddening values for the
Milky Way and LMC components separately. for the Milky Way E(B-V) = 0.070, while for the LMC E(B-V) = 0.075. The total reddening is E(B-V) = \NaEbv, in good agreement with the value obtained from the spectral analysis.

\section{Results} \label{sec:results}

To obtain the final solution we combine the results from the Monte Carlo analysis of the light and RV curves corrected for the presence of extra components with those from the NLTE spectroscopic analysis. The final results are summarized in Table~\ref{tab:properties}, where we give the physical parameters of the components together with the orbital properties of the inner system. Note that the temperature difference between the primary and secondary is determined much better than their individual temperatures.
From the binary model $\Delta T = 510 \pm 40$, which is almost identical to the one obtained from the spectroscopic analysis ($\Delta T = 500 \pm 100$).

\begin{figure*}
    \begin{center}
      \includegraphics[width=0.4\textwidth]{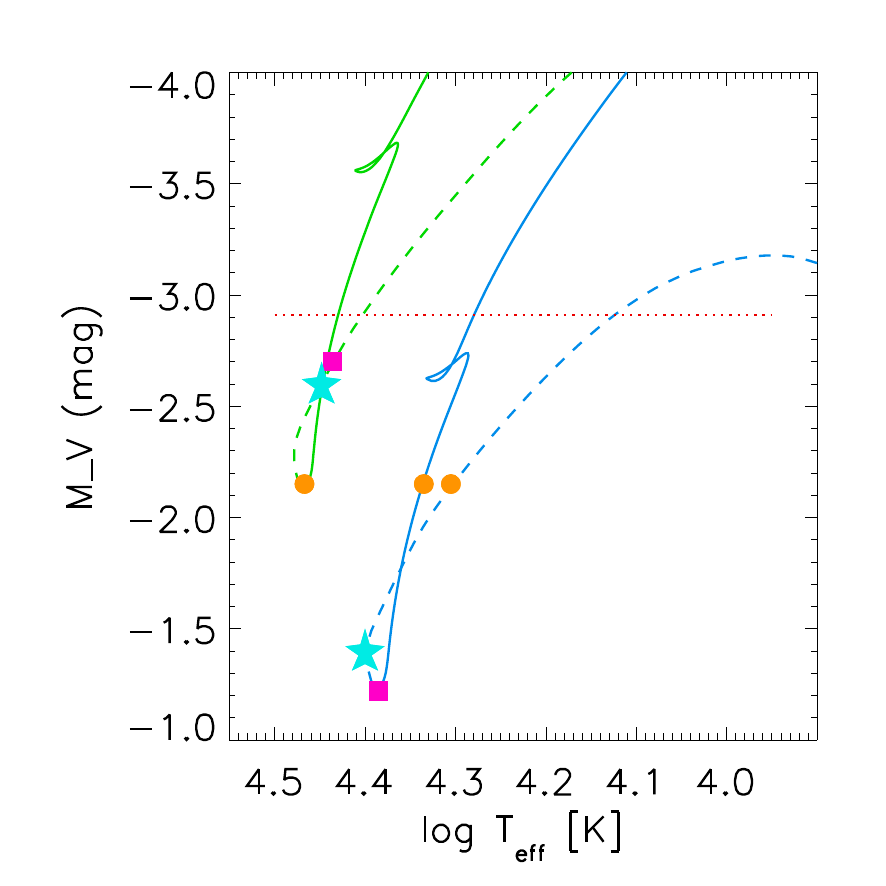}
      \includegraphics[width=0.4\textwidth]{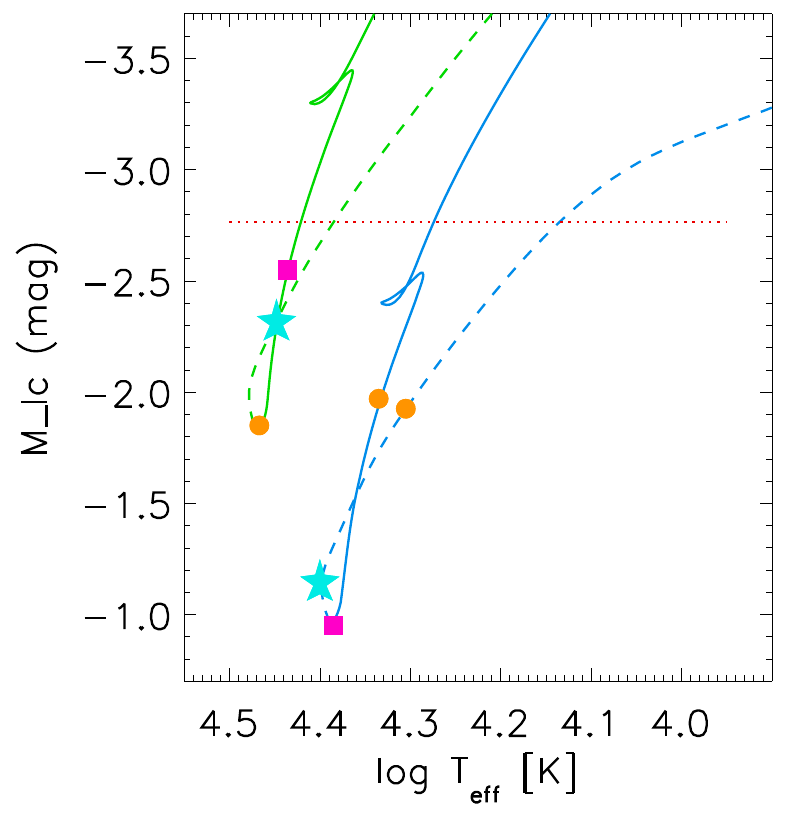}
      \includegraphics[width=0.4\textwidth]{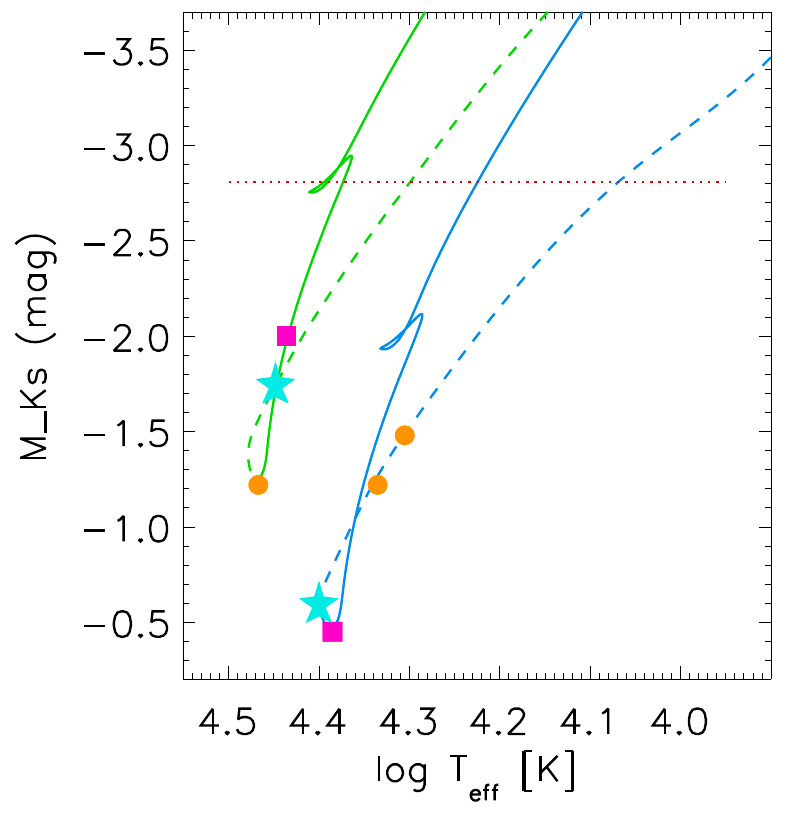}
    \end{center}
    \caption{Evolutionary constraints on the 3rd and 4th objects. We show MESA evolutionary tracks with ZAMS masses of 7.4 (blue) and 11 (green) M$_{\odot}$ corresponding to the minimum masses of the 4th and 3rd objects, respectively. Absolute magnitudes in the V-band (upper left), I$_C$ (upper right), and K$_s$ (lower panel) are plotted versus effective temperature. The dashed part of the tracks corresponds to the evolution towards the ZAMS. The horizontal dotted red line corresponds to the de-reddened total absolute magnitude of the combined third and fourth objects. The orange circles, pink squares, and cyan stars on the tracks are defined and discussed in the text.}
    \label{fig:3evo}
\end{figure*}

The distance modulus m-M = 18.59 $\pm$ 0.06 mag obtained through our spectral analysis and de-reddening can be
compared with the value of 18.45$\pm$0.02 mag derived from the one percent precision distance to the LMC determined by
\cite{Pietrzynski2019} and a small geometrical correction accounting for the location of the system within the LMC using the geometrical LMC model by  \cite{vandermarel2014}. Being 2 $\sigma$  larger, it just agrees within the error margins. It is interesting to note that for BLMC-02 analyzed in Paper II we also found a spectroscopic distance modulus 0.05 mag larger. This may hint at a small systematic effect caused by the use of our non-LTE model atmospheres, but we will need the analysis of additional objects to investigate this further.

With our determination of effective temperatures and luminosities, we can discuss the evolutionary status of the O-star binary system in the Hertzsprung-Russell diagram. In Fig.~\ref{fig:HRD} we show MESA evolutionary tracks \citep{Choi2016, Dotter2016} for 18, 19, 20 M$_{\odot}$ and the location of the primary and secondary in the HRD. From the MESA grid of evolutionary models, we have selected tracks with a metallicity [Z] = -0.5 and without rotation. We have selected the latter because the rotational velocities are relatively small compared to the alternative in the MESA grid with $v_{rot}/v_{crit} = 0.4$). According to the tracks, the binary components have a consistent age of 6.3 million years. The tracks indicate masses of 20 and 18 M$_{\odot}$, slightly higher than obtained with our analysis but at the margin of the uncertainties. 

\begin{table*}
    \begin{center}
    \caption{Physical parameters and other properties of the inner binary (A)}    \label{tab:properties}
    \begin{tabular}{l cc }
    \hline \hline
    Parameter [unit]            & Primary (Aa)                  & Secondary (Ab)   \\
    \hline
    mass [$M_\odot$]      & \VyE{\mPri}{\mPriErr}         & \VyE{\mSec}{\mSecErr}    \\ 
    radius [$R_\odot$]    & \VyE{\rPri}{\rPriErr}         & \VyE{\rSec}{\rSecErr}   \\
    $\log g$ [cgs]        & \VyE{\loggPri}{\loggPriErr}   &  \VyE{\loggSec}{\loggSecErr}    \\ 
    temperature [K]       & \VyE{\teffPri}{\teffPriErr}   & \VyE{\teffSec}{\teffSecErr}    \\ 
    $(T_2 - T_1)_{spec}$ {[K]}  & \multicolumn{2}{c}{$500 \pm 100$} \\
    $(T_2 / T_1)_{phot}$  & \multicolumn{2}{c}{$0.9851 \pm 0.0011$} \\
    $\log L$ [$L_\odot$]  & \VyE{\logLPri}{0.010}         & \VyE{\logLSec}{0.003}   \\
    $j_{21,V} = F_{2}/F_{1} (V)$          & \multicolumn{2}{c}{0.971 $\pm$ 0.002}              \\
    $V$ [mag]             & \VyE{\mVPri}{\mVPriErr}       & \VyE{\mVSec}{\mVSecErr}     \\
     $I_\mathrm{C}$ [mag]          & \VyE{\mIPri}{\mIPriErr}   & \VyE{\mISec}{\mISecErr}    \\
    $K_\mathrm{S}$ [mag]           & \VyE{\mKPri}{\mKPriErr}   & \VyE{\mKSec}{\mKSecErr}      \\
    $v_{rot}$ [$km\,s^{-1}$]             & \vrotBFPri                & \vrotBFSec           \\
    E(B-V) [mag]                 & \multicolumn{2}{c}{ {\avgEbvErr} }        \\ 
    orbital period, $P$ [days]        & \multicolumn{2}{c}{ \VyE{\per}{\perErr} }     \\
    semi-major axis [$R_\odot$]  & \multicolumn{2}{c}{\VyE{\sma}{\smaErr}}  \\
    orbital inclination, $i$                  & \multicolumn{2}{c}{ \VyE{\incl}{\inclErr}  }   \\
    eccentricity, $e$                 & \multicolumn{2}{c}{ \VyE{\ecc}{\eccErr} }     \\
    argument of periastron, $\omega$ [rad]               & \multicolumn{2}{c}{ \VyE{\aop}{\aopErr} }   \\
    apsidal motion, $\dot{\omega}$ [rad/d]       & \multicolumn{2}{c}{ \VyE{\daop}{\daopErr} }  \\
    orbital semiamplitude, $K$ [km s$^{-1}$]   & 228.5 $\pm$ 1.3   &  248.7 $\pm$ 1.1   \\
    systemic velocity, $\gamma$ [km s$^{-1}$]  &   \multicolumn{2}{c}{  \VyE{\vga}{\vgaErr}}          \\
    mass ratio, $q = m_2/m_1$   &  \multicolumn{2}{c}{ \VyE{\q}{\qErr}  }  \\
    \hline
    \end{tabular}
    \end{center}
    \tablecomments{Passband magnitudes are observed (reddened) values. For $\omega$ any value is possible within 3$\sigma$.}
\end{table*}

Having determined the masses of the inner system components we can also estimate them for the additional components. Assuming an orbital inclination of 90 degrees (outer orbit seen edge-on) from the spectroscopic orbit of the A system, the minimum mass of the tertiary is estimated to be about $M_B \sim 10.7 M_\odot$. With this limit at hand, and having the orbital semiamplitude of the A+B system derived from the detected LTTE ($K_{AB} = 8.6 \pm 0.4$), we obtain the minimum mass of the fourth component of $M_C \sim 7.3 M_\odot$.
Moreover, as we know that these objects provide about 20\% of the total system light and are not visible in the spectra, the individual components should be less massive than the primary and secondary components, individually. This puts a conservative upper limit to the mass of about $17.5 M_\odot$ (the mass of the secondary) for a single star. From the fact that the more inclined the outer orbit is the more massive tertiary we would obtain, we can also constrain the inclination of the A+B system, $i_{A+B} \gtrsim 42^\circ$, assuming B is a single star. With the same assumption, we expect $M_B$ to be close to the lower limit and to have an inclination similar to that of the inner system, although the orbits do not necessarily have to be co-planar.

We can use the additional light found in our light curve analysis of section~\ref{sec:finalmodel} for an attempt to further constrain the nature of the 3rd and 4th objects. With the magnitude corrections given above, we obtain V = 16.05$\pm$0.03 mag, V - I$_c$ = 0.06$\pm$0.02 mag and V - K$_s$ = 0.35$\pm0.04$ mag. With E(B-V) = 0.16 mag and R$_V$ = 3.2 (see previous section) we can correct for extinction and calculate intrinsic colors and magnitude: V$_0$ = 15.54$\pm$0.08 mag,   (V - I$_c$)$_0$ = -0.15$\pm$0.10 mag and (V - K$_s$)$_0$ = -0.10$\pm0.10$ mag.

 With a distance modulus of 18.45 (see above) we then obtain absolute magnitudes M$_V$ = -2.91$\pm$0.08 mag, M$_{I_C}$ = -2.76$\pm$0.06 mag and M$_{K_s}$ = -2.80$\pm$0.05 mag. This allows us to compare the joint magnitudes of the objects with MESA evolutionary tracks of 7.4 and 11 M$_{\odot}$, corresponding to the minimum masses of the fourth and third object, respectively. This is done in Fig.~\ref{fig:3evo} where we show absolute magnitudes in V, I$_C$, and K$_s$ versus effective temperature. The MESA tracks are from the same evolutionary grid as used in Fig.~\ref{fig:HRD}. They include pre-main sequence evolution towards the zero-age main sequence (ZAMS) and post-main sequence evolution away from ZAMS. The observed absolute magnitudes of the combined third and fourth objects are also shown in each panel.

Fig.~\ref{fig:3evo} indicates that only a small range of effective temperature of the tracks is compatible with the observed magnitudes. From the V-band we see immediately that the effective temperature of the 3rd object (represented by the track with $11 M_\odot$) is tightly constrained to an evolutionary phase very close to the ZAMS with log T$_{eff} \ge$ 4.4. This also restricts the evolutionary status of the 4th object (the track with $7.4 M_\odot$), because the combined brightness of both objects needs to match the observed magnitude shown as the red dotted line. At the minimum brightness of the 3rd object, the 4th object must have a very similar magnitude as indicated by the orange circles on the tracks. The maximum possible brightness of the 3rd object is constrained through the minimum V-band magnitude of the 4th object on the $7.4 M_\odot$ track. The pink squares on each track define the corresponding evolutionary phases and magnitudes. The range between the pink squares and the orange circles on both tracks describes the evolutionary phase during which the joint brightness of both objects matches the V-band observation. The cyan stars provide an example within this range.

 We then use exactly the same evolutionary phases represented by the orange circles, pink squares, and cyan stars in the plot of M$_{I_C}$ in Fig.~\ref{fig:3evo}. Combining the corresponding magnitudes on the two tracks we find that they also agree with the observation within the error margins. However, we encounter a severe discrepancy with the observations for  M$_{K_s}$ as shown in Fig.~\ref{fig:3evo}. The combined magnitudes of the orange circles, cyan stars, and pink squares, respectively, are 0.6 to 0.75 magnitudes too faint. This may indicate an IR excess of these objects, for instance, caused by a circumstellar disk or an even more complex structure of the system with a possible multiplicity of the B and C components.

In a similar analysis, we also found that evolutionary track for a star of 16 M$_\odot$ would lie entirely above the red dotted line in the V-band panel of Fig.~\ref{fig:3evo} (passing slightly below it for the I-band), i.e. the star at any evolutionary stage would be brighter than the additional light we found. This puts a stronger (than the mass of the secondary) mass limit of 16 M$_\odot$ on the 3rd and 4th component (or individual stars they consist of in the case any of them is a binary itself).

\section{Conclusions}\label{sec:conclusions}

From our multi-method analysis, the BLMC-03 system resulted to be much more complex than expected.
We found strong evidence for the existence of two extra components orbiting around the studied binary system. Each of these two extra components may be a binary itself but we have no direct evidence for that, apart from a small hint from the additional light in the Ks-band, which may suggest more than 2 but cooler and redder stars in the system.

This (at least) quadruple system belongs to the less common group of 2+1+1 systems (2+2 are more frequent, see e.g. \citealt{tokovinin:2014}). The period ratios are $P_{A+B}/P_{A} = 60.7$ and $P_{AB+C}/P_{A+B} = 13$. The system is thus relatively tight for a 3-tier hierarchical system.

One may question if such a complex system can be used for our final goal of the calibration of the SBC relation. However, we have full control over the additional light in the system and take into account the effects of additional components on the measured quantities. As a result, the accuracy and precision of determined parameters are comparable to those for our other systems. One may also argue that it is better to have a binary as a part of a multiple system with detected additional components than a binary with such components not yet detected. This also points to the importance of high-quality data and a detailed study of any extragalactic binary system that will be used in the future for distance measurement.

\section{Acknowledgments}

We thank Dr. Nidia Morrell, Dr. Philip Massey and Dr. Kathryn Neugent for sharing with us the MagE spectra of BLMC-03. We also thank Nidia Morrell for acquiring a few additional recent MIKE spectra that helped constrain the outer orbit.
The research leading to these results has received funding from the European Research Council (ERC) under the European Union's Horizon 2020 research and innovation programme under grant agreement No 951549 (project UniverScale) and from the National Science Center, Poland grant BEETHOVEN UMO-2018/31/G/ST9/03050. 
RPK has been supported by the Munich Excellence Cluster Origins funded by the Deutsche Forschungsgemeinschaft (DFG, German Research Foundation) under Germany's Excellence Strategy EXC-2094 390783311.
M.T., B.P., and G.P. received support from the Munich Institute for Astro-, Particle- and BioPhysics (MIAPbP) which is also funded by the Deutsche Forschungsgemeinschaft (DFG, German Research Foundation) under Germany's Excellence Strategy EXC-2094 390783311.
B.P. acknowledges funding from the Polish National Science Center grant SONATA BIS 2020/38/E/ST9/00486. 
We also acknowledge support from the Polish Ministry of Science and Higher Education grant DIR-WSIB.92.2.2024. The research was based on data collected under the Polish-French Marie Skłodowska-Curie and Pierre Curie Science Prize awarded by the Foundation for Polish Science.

This research is based on observations collected at the Las Campanas Observatory and the European Southern Observatory under ESO programs: 098.D-0263(A), 0100.D-0339(A), 0102.D-0469 and 102.D-0590 (B). 

This research has made use of NASA's Astrophysics Data System Service.

\vspace{5mm}
\facilities{VLT:Kueyen (UVES), Magellan:Clay (MIKE and MagE)}

\software{\texttt{ESO Reflex} \citep[][\url{http://www.eso.org/sci/software/esoreflex/}]{freudling:2013}, 
\texttt{PHOEBE} \citep[][\url{http://phoebe-project.org/1.0}]{prsa:2005},\\
\texttt{RaveSpan} \citep[][\url{https://users.camk.edu.pl/pilecki/ravespan/index.php}]{pilecki:2017},
\texttt{FASTWIND} \citep[][]{Puls2005, Rivero2012}
}

\bibliographystyle{aasjournal}
\bibliography{taormina2024_paper3.bib}

\end{document}